# Deep Reinforcement Learning for Long-Short Portfolio Optimization

Gang Huang[1]*, Xiaohua Zhou[1] and Qingyang Song[2]
[1]Department of Applied Economics, Chongqing University, 174 Shazheng Street, Chongqing, 400044, China.
[2]Department of Technology Economics and Management, Chongqing University, 174 Shazheng Street, Chongqing, 400044, China.

*Corresponding author's E-mail: huanggangvoyager@163.com;

**Abstract:** The rapid advancement of artificial intelligence technology enables data-driven methods to effectively overcome limitations inherent in traditional portfolio optimization approaches. While traditional portfolio optimization models predominantly employ long-only strategies and deliberately exclude highly correlated assets to diversify risk, the introduction of short-selling mechanisms facilitates low-risk arbitrage through hedging correlated assets. This research develops a portfolio management framework based on deep reinforcement learning (DRL) with a short-selling mechanism designed to comply with actual trading rules, seeking to generate excess returns in China's A-share market. Key innovations include: (1) Implementation of a comprehensive short-selling mechanism for continuous trading environments that accurately captures the dynamic transaction process across consecutive trading periods; (2) Development of an integrated long-short portfolio optimization framework that combines specialized deep neural network architectures for multi-dimensional financial time series processing with average Sharpe ratio reward functions and advanced DRL algorithms. Empirical testing reveals that the DRL model with short-selling capabilities demonstrates superior optimization performance, generating consistent positive returns across both back-testing periods. Relative to conventional optimization approaches, the proposed model delivers enhanced risk-adjusted returns while substantially reducing maximum drawdown. The asset allocation profile indicates the DRL model establishes a resilient investment strategy by systematically avoiding historically underperforming assets and maintaining balanced capital distribution, thereby strengthening the portfolio's defensive capabilities. This study contributes to portfolio theory advancement while providing novel methodological approaches for quantitative investment implementation.

## 1 INTRODUCTION

Exploration of scientific investment decision-making contributes both to theoretical researchers' understanding of securities market microstructure and to practitioners' access to improved decision frameworks and methodologies. Extensive empirical evidence indicates that stock markets in developed and developing economies consistently deviate from efficiency, with market anomalies being pervasive. Within this context, traditional portfolio optimization theories, despite their contributions, exhibit significant limitations.

Traditional portfolio optimization models predominantly employ long-only mechanisms, strategically excluding highly correlated assets to achieve risk diversification. However, the incorporation of short-selling mechanisms enables low-risk arbitrage through the hedging of correlated assets, substantially enhancing optimization capabilities. Critically, conventional models typically conceptualize optimization as a static problem, neglecting the dynamic characteristics inherent in asset behavior during actual trading processes. Even models that incorporate short-selling mechanisms often fail to adopt a holistic portfolio perspective, resulting in distorted return calculations. This distortion primarily originates from the constraints of traditional asset weight definitions, which require the sum of all asset weights to equal 1 (i.e., $\sum_i^m \omega_i = 1$). While appropriate for long-only positions, this restrictive definition produces return calculations that materially diverge from actual trading outcomes when short positions are involved.

Recent advances in artificial intelligence have enabled data-driven approaches to effectively transcend the constraints of traditional methods. The reinforcement learning (RL) paradigm, exemplified by China's DeepSeek-R1 technology, has demonstrated transformative capabilities in complex reasoning tasks, highlighting RL's broad applicability in dynamic modeling of complex systems. Deep reinforcement learning-based portfolio optimization eliminates the need for predetermined variable relationships, instead deriving investment strategies through iterative agent-environment interactions. This methodology accommodates market non-linearities while more accurately representing trading returns through refined short-selling weight calculation mechanisms.

This research develops a portfolio management framework based on deep reinforcement learning (DRL), implementing a short-selling mechanism aligned with actual trading regulations to identify effective strategies for generating excess returns in the A-share market. The study advances portfolio theory while introducing innovative approaches for quantitative investment implementation.

## 2 LITERATURE REVIEW

Markowitz's [1] modern portfolio theory established a fundamental framework for contemporary investment decision-making. This seminal work introduced statistical methodology to portfolio selection, quantifying the return-risk relationship through a mean-variance paradigm and transforming investment decisions from subjective assessments to quantitative analysis. Building upon Markowitz's mean-variance framework, researchers have developed numerous refinements to address its limitations. The Black-Litterman model [2, 3] employs Bayesian methodology to synthesize subjective market views with equilibrium assumptions, mitigating parameter estimation errors. Conditional Value-at-Risk (CVaR) [4] models focus specifically on downside risk, providing more accurate representation of portfolio tail risk characteristics. Risk parity [5] and hierarchical risk parity [6] approaches reformulate asset allocation from a risk distribution perspective, reducing the sensitivity to expected return estimates that plague traditional methodologies. Concurrent with computational advances, non-parametric techniques based on Data Envelopment Analysis (DEA) [7] and heuristic optimization algorithms including simulated annealing [8] and genetic algorithms [9] have expanded the portfolio optimization toolkit, offering effective solutions for high-dimensional optimization challenges. While these theoretical and methodological developments have substantially enriched portfolio optimization frameworks, they predominantly operate within long-only constraints and inadequately address the implications of short-selling on portfolio construction, thereby restricting investment strategy flexibility.

As financial markets have evolved, scholars have increasingly recognized the limitations of traditional portfolio optimization theories regarding short-selling mechanisms and have conducted extensive research addressing these deficiencies. From a theoretical perspective, Ross's [10] Arbitrage Pricing Theory (APT), introduced in 1976, while not explicitly focused on short-selling, established the foundational arbitrage portfolio construction framework that would inform subsequent short-selling strategy research. This theory transcended single-factor pricing model constraints by explaining asset returns through multiple risk factor linear combinations, providing theoretical support for risk management within short-selling strategies.

In strategy development, Jacobs and Levy [11] conducted systematic analyses of trading approaches incorporating short positions, including market-neutral and hedging strategies. Their work encompassed both theoretical articulation of short-selling risk-return characteristics and empirical validation across diverse market conditions. Lo and Patel [12] advanced portfolio construction methodology by developing the "130/30" strategy, which enhances portfolio risk-return profiles through an optimization framework combining 130% long exposure with 30% short exposure. Gatev et al. [13] introduced pairs trading methodology from a correlation perspective, challenging the traditional portfolio theory practice of avoiding highly correlated assets and pioneering arbitrage techniques that specifically leverage these high correlations.

At the model optimization level, Jacobs et al. [14] developed computational methodology for efficiently deriving mean-variance efficient frontiers with short-selling constraints using factor models and scenario analysis. While this approach comprehensively addresses short position modeling and allocation frameworks, it requires further refinement regarding transaction cost incorporation and market dynamic adaptation. Dhingra et al. [15] extended short-selling strategies to the mean-expected risk value framework, constructing portfolio models that integrate risk control with return optimization. Despite limitations including parameter sensitivity issues and inadequate transaction cost treatment, this research establishes new methodological approaches for applying econometric models to short-selling implementations.

While these studies have advanced short-selling strategy development across multiple dimensions, significant limitations persist in modeling short-selling mechanisms within continuous trading environments. Current methodologies predominantly conceptualize asset allocation as a static process; even studies incorporating time series elements frequently neglect portfolio dynamic adjustment characteristics during continuous trading. Particularly when accounting for transaction costs and market dynamics, comprehensive theoretical frameworks for constructing dynamic weight allocation methods compliant with practical trading regulations in short-enabled portfolios remain underdeveloped. This theoretical gap constrains both academic advancement and practical strategy implementation. As capital markets continue to evolve and investment strategies become increasingly sophisticated, the development of short-selling mechanisms aligned with actual trading processes has become critically important.

Deep reinforcement learning experienced a significant breakthrough in 2015 when Mnih et al. [16] successfully integrated deep learning with reinforcement learning methodologies, enabling end-to-end learning for Atari games through the Deep Q-Network architecture. This advancement demonstrated deep reinforcement learning's efficacy in sequential decision-making: agents develop optimal long-term strategies through environmental interaction. This capability holds particular relevance for portfolio optimization, as investment decisions fundamentally represent multi-period dynamic optimization problems. Deep reinforcement learning frameworks can incorporate market states, transaction costs, and other relevant factors, developing optimal trading execution strategies through iterative learning processes. This approach enables end-to-end optimization from market data to trading decisions, more accurately reflecting actual trading dynamics than conventional single-period optimization methodologies.

Before this deep reinforcement learning breakthrough, asset management applications of reinforcement learning primarily employed Q-learning and Policy Gradient (PG) algorithms. Moody [17] developed a PG-based single-asset management model, with subsequent derivative models predominantly focusing on single-risk asset management or fixed investment decision frameworks, exemplified by Dempster et al.'s [18] automated foreign exchange trading system and Zhang et al.'s [19] asset management framework. Similarly, Q-learning applications proposed by Ralph Neuneier [20], Gao et al. [21], and Lee et al. [22] remained confined to single-asset management contexts. With recent breakthroughs in Deep Reinforcement Learning (DRL) technology, single-asset management models have demonstrated significant advancements in feature extraction capabilities, policy adaptability, and trading performance [23, 24, 25, 26, 27]. These developments indicate that DRL technology is driving single-asset management toward more intelligent and efficient methodologies.

Research on multi-asset portfolio optimization using deep reinforcement learning offers substantially greater practical relevance and implementation value compared to single-asset prediction and trading strategy studies. Recent years have witnessed extensive scholarly exploration in this domain, yielding significant methodological advancements. Jiang et al. [28] developed a DRL portfolio optimization framework for cryptocurrency assets, demonstrating DRL's potential applications within financial contexts. Wu et al. [29] constructed a portfolio management framework based on DRL that resolved constraint challenges in continuous action spaces through enhanced proximal policy optimization algorithms. Yang [30] introduced the TC-MAC algorithm, which employs graph neural networks (GNN) to capture inter-asset dynamic relationships for portfolio optimization. However, this latter research contains notable methodological limitations: despite defining a reward function based on \$10,000 initial capital, training rewards converged to approximately \$3,000 maximum, indicating potential significant capital deterioration under the proposed strategy. Furthermore, the model disregards intermediate transaction cost dynamics in cost definition and employs questionable transaction cost multiplier factor methodology in wealth calculation. These theoretical framework deficiencies substantially undermine the reliability of the model's results.

The research described above predominantly addresses long-only optimization strategies for multiple assets; however, in practical financial markets, short-selling mechanisms offer investors enhanced trading flexibility and risk management capabilities. Research applying deep reinforcement learning to portfolio optimization incorporating short-selling remains comparatively underdeveloped. Within the limited existing literature on long-short strategies,

Almahdi and Yang [31] developed an adaptive trading system based on Recurrent Reinforcement Learning (RRL). This framework exhibits two principal limitations in long-short portfolio optimization: it lacks precise formulations for weight constraint conditions in non-equal-weighted portfolios, and it reduces short-selling to discrete trading signals with definitional inconsistencies between weights and signals, compromising its applicability in continuous trading environments. Sun et al. [32] proposed an integrated approach combining DRL with the Black-Litterman model for portfolio optimization, implementing long-short strategies through dynamic inter-asset correlation analysis. However, this model contains two significant methodological constraints: it restricts weight allocation through exogenously determined long and short position parameters, limiting the DRL model's capacity for autonomous optimal weight discovery and diverging from data-driven optimization principles; additionally, it employs traditional sum-to-one weight constraints rather than absolute-sum weight constraints, potentially resulting in systematic overestimation of portfolio returns in long-short strategy implementations.

A comprehensive analysis of extant literature indicates that traditional portfolio optimization theories inadequately address both the dynamic characteristics of trading processes and short position optimization. While reinforcement learning-based portfolio optimization methodologies have emerged, they continue to demonstrate incomplete modeling of short-selling costs and constraints, frequently incorporating exogenously determined trading restrictions that conflict with data-driven optimization principles. To address these limitations, this research contributes the following innovations:

First, we develop a comprehensive short-selling mechanism for continuous trading environments that accurately models weight evolution processes in long-short portfolios by implementing a constraint where the sum of absolute asset weights equals one ($\sum_{i=0}^{m}|\omega_i| = 1$). This formulation enables the model to dynamically recalculate short position weights in response to market price fluctuations and accumulate corresponding transaction costs. The mechanism ensures precise short position valuation while establishing an operationally viable theoretical foundation for practical trading implementation.

Second, we construct an integrated long-short portfolio optimization framework that employs random sampling strategies during training and synthesizes deep neural network architectures specialized for multi-dimensional financial time series processing with average Sharpe ratio reward functions and advanced deep reinforcement learning algorithms. This integration enables unified optimization across both long and short strategies, substantially enhancing overall model performance.

## 3 DRL MODEL CONFIGURATION

DRL represents a dynamic optimization methodology conforming to Markov Decision Process (MDP) principles. Portfolio asset trading processes can be effectively conceptualized as an MDP, with the complete sequence from account initiation to trading conclusion forming a trajectory $\tau = (S_0, A_0, R_1, S_1, A_{1,} R_2, S_2, A_{2,} R_3, \cdots)$. This trajectory, commonly termed an episode, is particularly amenable to DRL theoretical modeling approaches. Following established DRL modeling protocols, this research frames a portfolio manager as an agent, systematically defines the state (environment), action, and reward components, and subsequently implements a selected DRL algorithm with appropriate deep neural network architecture to optimize portfolio performance.

3.1 State Space Configuration

The state $S_t = X_t$, where the price tensor $X_t$ comprises multiple data features: the daily opening price $V_t^{(op)}$, lowest price $V_t^{(lo)}$, highest price $V_t^{(hi)}$, and closing price $V_t^{(cl)}$. Fig.1 illustrates this data structure.

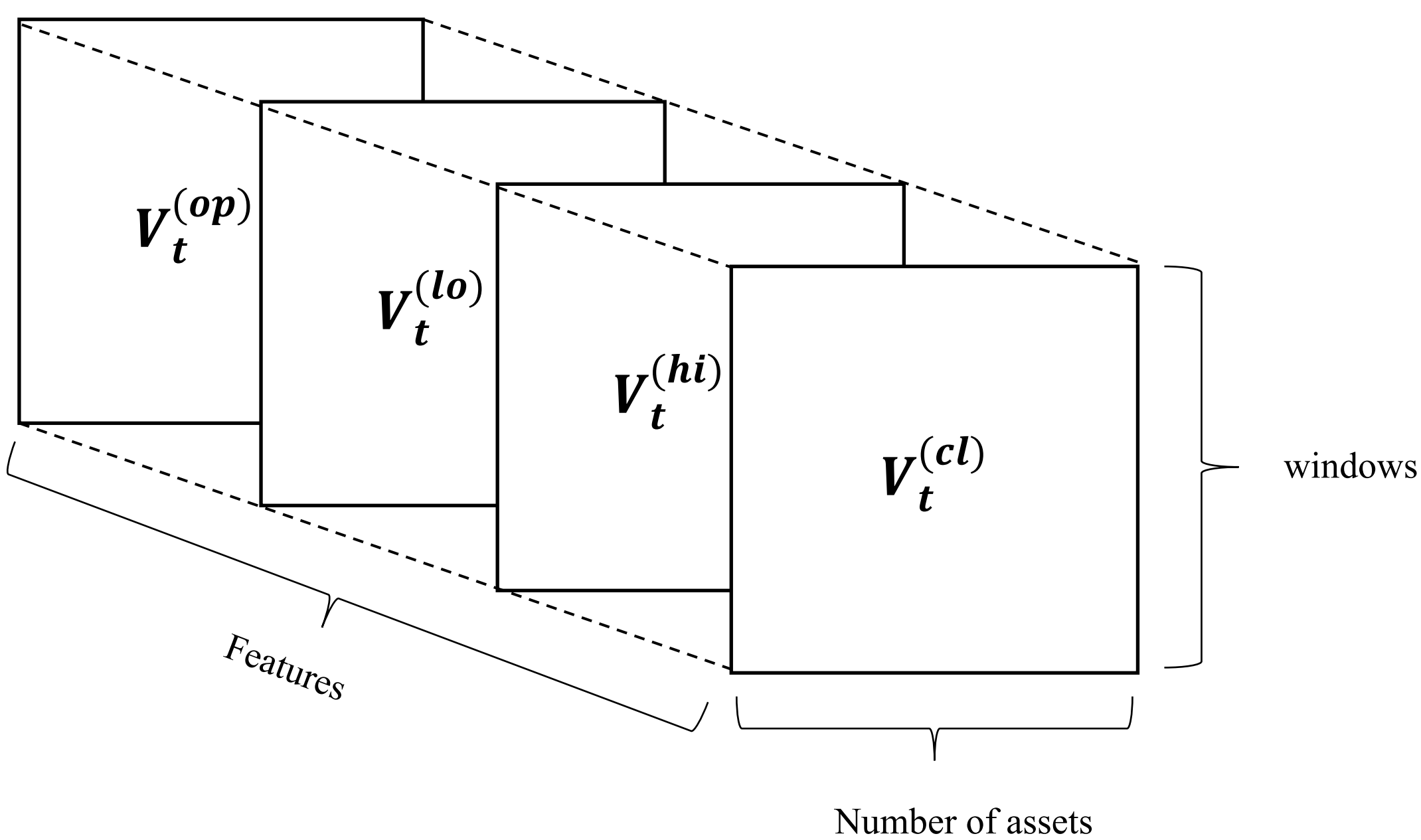

**Fig.1** Data structure of state $X_t$

The tensor calculation formulas are as follows:

$$
\begin{aligned}
\boldsymbol{V}_t^{(op)} &= [\boldsymbol{v}_{t-n+1}^{(op)} \oslash \boldsymbol{v}_t | \boldsymbol{v}_{t-n+2}^{(op)} \oslash \boldsymbol{v}_t | \cdots | \boldsymbol{v}_{t-1}^{(op)} \oslash \boldsymbol{v}_t | \boldsymbol{v}_t^{(op)} \oslash \boldsymbol{v}_t] \\
\boldsymbol{V}_t^{(lo)} &= [\boldsymbol{v}_{t-n+1}^{(lo)} \oslash \boldsymbol{v}_t | \boldsymbol{v}_{t-n+2}^{(lo)} \oslash \boldsymbol{v}_t | \cdots | \boldsymbol{v}_{t-1}^{(lo)} \oslash \boldsymbol{v}_t | \boldsymbol{v}_t^{(lo)} \oslash \boldsymbol{v}_t] \\
\boldsymbol{V}_t^{(hi)} &= [\boldsymbol{v}_{t-n+1}^{(hi)} \oslash \boldsymbol{v}_t | \boldsymbol{v}_{t-n+2}^{(hi)} \oslash \boldsymbol{v}_t | \cdots | \boldsymbol{v}_{t-1}^{(hi)} \oslash \boldsymbol{v}_t | \boldsymbol{v}_t^{(hi)} \oslash \boldsymbol{v}_t] \\
\boldsymbol{V}_t^{(cl)} &= [\boldsymbol{v}_{t-n+1} \oslash \boldsymbol{v}_t | \boldsymbol{v}_{t-n+2} \oslash \boldsymbol{v}_t | \cdots | \boldsymbol{v}_{t-1} \oslash \boldsymbol{v}_t | \boldsymbol{1}]
\end{aligned} \tag{1}
$$

In formula (1), t denotes the t-th trading period, and the symbol $\oslash$ represents element-wise division, where each element in the vectors on both sides is divided by its corresponding positional element. The lowercase letter $\boldsymbol{v}_t$ denotes the closing price on the t-th trading day. Each element in the price tensor $\boldsymbol{X}_t$ is divided by the closing price $\boldsymbol{v}_t$, effectively normalizing the price data.

3.2 Action Space Configuration

This paper defines the weights of assets in the portfolio (Asset Weights) as actions within the DRL framework, where $\boldsymbol{Actions} = \boldsymbol{Weights}$. Consequently, the DRL Action is formulated as:

$$\boldsymbol{W_t} = \left(\boldsymbol{\omega_{0,t}}, \boldsymbol{\omega_{1,t}}, \boldsymbol{\omega_{2,t}}, \cdots, \boldsymbol{\omega_{m,t}}\right) \tag{2}$$

This section examines the short-selling mechanism in stock markets. When engaging in short-selling, investors must acquire the target assets through designated channels (primarily via the securities lending system in China's A-share market). The process follows a specific sequence: investors initially borrow stocks from lenders (who simultaneously require appropriate collateral or collateral assets), then sell these stocks at prevailing market prices. When stock prices reach the investor's target level, the investor repurchases an equivalent quantity of shares to return to the lender, reclaims their collateral assets, and realizes profit from the differential between selling and purchasing prices.

For analytical simplicity, this model assumes zero holding costs for borrowed stocks and excludes any dividends or distributions accrued during the holding period. The short-selling leverage ratio is fixed at 1:1, requiring that the total value of collateral assets equals the current market value of borrowed stocks (calculated at current prices), with no additional leverage permitted.

Within the portfolio framework, negative weights denote short positions, which yield positive returns during price declines. Conversely, positive weights denote long positions, which generate losses during price declines. Long positions yield positive returns exclusively during price appreciation, functioning as the precise inverse of short-selling strategies.

Therefore, the action vector $\boldsymbol{W_t}$ is constrained to the range [-1,1], with all portfolio asset weights subject to the following constraint:

$$\sum_{i=0}^{m} |\omega_i| = 1 \tag{3}$$

The weights of assets in the portfolio fundamentally represent the proportion of each asset's current market value relative to the total portfolio market value. Consistent with short-selling mechanics, negative weights indicate short positions, while positive weights denote long positions. Within the DRL action space framework, the cross-sectional sum of absolute values of all asset weights $\boldsymbol{\omega_i}$ in the portfolio strictly equals 1.

In this framework, $\boldsymbol{\omega_0}$ represents cash assets, which cannot be subject to short-selling, thus constraining the domain of $\boldsymbol{\omega_0}$ to [0,1]. The following example illustrates the weight calculation methodology for short positions: if at time t, we utilize assets worth 200,000 yuan as collateral to borrow stocks of asset a with equivalent market value for short-selling, while the combined market value of other portfolio assets equals 800,000 yuan, then the weight of asset a at time t is calculated as $\boldsymbol{\omega_a} = \frac{-20}{80 + |-20|} = -0.2$.

The portfolio weights are initialized as $\boldsymbol{W_0} = (1,0,\cdots,0)^T$, indicating that prior to trading initiation, the portfolio consists exclusively of cash assets with zero allocation to other assets. In this paper, the complete portfolio composition encompasses 1 cash asset + m risky assets.

### 3.3 Other Elements Derivation and Reward Function Setting

The relative price vector $\boldsymbol{Y_t}$ for trading in period t is defined as:

$$\boldsymbol{Y_t} \triangleq \boldsymbol{P_t} \oslash \boldsymbol{P_{t-1}} = \left(1, p_{1,t} / p_{1,t-1}, \cdots, p_{i,t} / p_{i,t-1}\right)^T \tag{4}$$

$\boldsymbol{P_t}$ represents the vector of closing prices for all assets in the portfolio at period t. For simplicity, we assume that the growth rate of cash remains 0 throughout the investment period; consequently, the first element in the relative price vector $\mathbf{Y_t}$ is consistently 1.

If $\rho_t$ represents the value of the portfolio in period t, ignoring transaction costs, then the value of the portfolio can be expressed as:

$$\rho_t = \rho_{t-1}\exp[(\mathbf{ln}\boldsymbol{Y_t}) \cdot \boldsymbol{W_{t-1}}] \tag{5}$$

where $\cdot$ represents the dot product of vectors, ln represents the natural logarithm, and exp represents the exponential function with the natural constant e as the base. The daily log return $\gamma_t$ of the portfolio is

$$\gamma_t = \ln(\rho_t \,/\, \rho_{t-1}) \tag{6}$$

Next, the mean $\bar{R}$ and standard deviation $Std(\gamma_t)$ of the daily log returns are calculated using the following formulas:

$$\bar{R} = \frac{1}{t_n}\sum_{t=1}^{t_n}\gamma_t \tag{7}$$

$$std(\gamma_t) = \sqrt{\sum_{t=1}^{t_n}(\gamma_t - \bar{R})^2 \Big/ t_n} \tag{8}$$

In equations 7 and 8, $t_n$ denotes the nth trading period, and $\gamma_t$ is calculated based on the closing price at the end of period t. When investors initially enter the market, they hold only cash assets. This initial trading point is defined as t=0 with $\gamma_0$=0.

The average annualized Sharpe ratio functions as the reward function, with the formula as follows:

$$\textbf{reward:}\ An_Sharpe_t = \sqrt{Freq} \cdot (\bar{R} - r_f) \Big/ Steps \cdot std(\gamma_t - r_f) \tag{9}$$

Freq denotes the number of trading days per year, set at 252 in this study. $r_f$ represents the risk-free asset return rate, corresponding to cash holdings with a return rate of 0 ($r_f$=0). Steps refers to the number of steps in a training episode, during which the agent executes one trading decision per step. This reward function demonstrates effective enhancement of model performance [33], with its maximization constituting the primary training objective.

## 3.4 Transaction Costs and Portfolio Returns

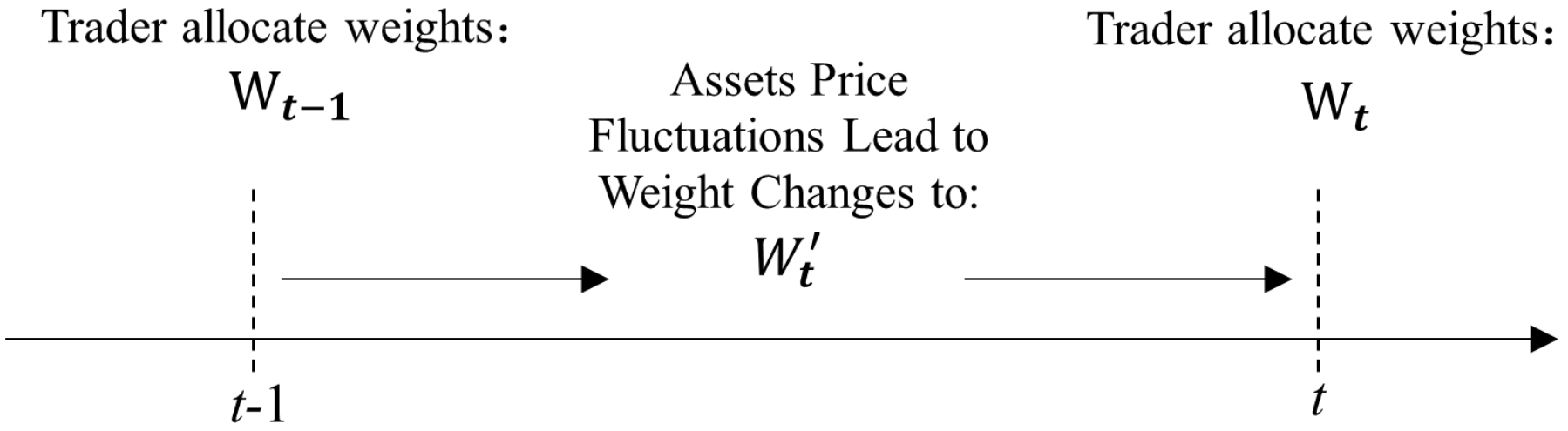

**Fig.2** Changes in Asset Weights

Portfolio transaction costs consist of fixed trading costs and variable trading costs [34], with the latter comprising market impact costs and opportunity costs. Given the challenges in quantifying variable costs, this analysis concentrates on fixed trading costs. The parameter $\mu_t$ denotes the transaction cost rate for an individual asset, representing commissions and other fixed transaction costs during the trading process. The portfolio weight vector at the commencement of period t is defined as $W_{t-1}$. Through natural market price fluctuations, this weight vector transforms to $W_t'$, according to the formula:

$$\boldsymbol{W_t'} = (\boldsymbol{Y_t} \odot \boldsymbol{W_{t-1}}) \,/\, (\boldsymbol{Y_t} \cdot |\boldsymbol{W_{t-1}}|) \tag{10}$$

The symbol $\odot$ denotes the Hadamard product, while the symbol $\cdot$ represents the dot product. The notation $|\ \ |$ signifies the application of the absolute value operation to each vector element, with the sum of these absolute values constrained to 1 (as referenced in equations 2 and 3). $\boldsymbol{W_t'}$ represents the portfolio weights resulting from natural asset price fluctuations during the interval between the completion of trading in period t-1 and the commencement of trading in period t. This weight evolution process is illustrated in Fig.2.

At the commencement of period t, the trader recalibrates the asset weights to $W_t$. This adjustment transforms the weight vector from $W_t'$ to $W_t$. The resulting transaction cost is expressed as:

$$C_t = \mu_t \left( \sum_{i=1}^{m} \left| \boldsymbol{\omega}_{i,t}^{'} - \omega_{i,t} \right| \right) \tag{11}$$

$\mu_t$ denotes the transaction cost rate for an individual asset in period t. Based on the empirical characteristics of the A-share market, $\mu_t$ is set at 0.0015, while $C_t$ represents the transaction cost rate for the entire portfolio. When accounting for transaction costs, the portfolio value is given by:

$$\rho_t = \rho_{t-1}(1 - C_t)\exp[(\mathbf{ln}\boldsymbol{Y_t}) \cdot \boldsymbol{W_{t-1}}] \tag{12}$$

Based on equations 6 and 12, the portfolio return rate incorporating transaction costs is expressed as:

$$\gamma_t = \mathbf{ln}(1 - C_t) + (\mathbf{ln}\boldsymbol{Y_t}) \cdot \boldsymbol{W_{t-1}} \tag{13}$$

where $\mathbf{ln}\boldsymbol{Y_t}$ denotes the vector of logarithmic returns for the individual assets in the portfolio. For notational simplicity, we define $\Gamma_t$ as the return vector derived from $\mathbf{ln}\boldsymbol{Y_t}$, yielding:

$$\begin{gathered} \gamma_t = \Gamma_t \cdot W_{t-1} + \mathbf{ln}(1 - C_t) \\ \text{subject to } \textstyle\sum_{i=0}^{m} \left| \omega_{i,t} \right| = 1 \end{gathered} \tag{14}$$

Equation 14 provides the portfolio return calculation formula within a dynamic trading framework that incorporates short-selling mechanisms. This study employs no financial leverage for any assets, maintaining a consistent leverage ratio of 1:1 across all securities.

## 4 DRL ALGORITHM AND NETWORK STRUCTURE

DRL algorithms refer to methodologies in which agents optimize decision policies through persistent interaction with their environment. Specifically, at each timestep, an agent assesses its current state, selects an optimal action from its action space, and incrementally refines its policy based on immediate rewards and environmental feedback to maximize cumulative returns. This study implements DRL algorithms using the Stable Baselines3 (SB3) reinforcement learning framework. DRL algorithms are broadly classified into on-policy and off-policy approaches, each demonstrating varied performance characteristics across different problem domains. During our experimental evaluation, we tested multiple algorithmic architectures and observed that off-policy algorithms generally required greater computational resources and exhibited slower convergence properties. Constrained by available hardware capabilities, we ultimately selected Proximal Policy Optimization (PPO) [35], an on-policy algorithm, as our primary DRL methodology.

### 4.1 PPO ALGORITHM

PPO represents a DRL algorithm built upon the PG framework. The core principle of PPO involves constraining the magnitude of policy updates to achieve efficient optimization while maintaining training stability. PPO exhibits notable advantages over traditional policy gradient methods in three key aspects: sample efficiency, parameter sensitivity, and convergence stability.

In the PPO algorithm formulation, where $\pi_\theta(a|s)$ is parameterized by θ, the expected objective function is defined as:

$$J(\theta) = \mathbb{E}_{\tau \sim \pi_\theta}\left[\sum_{t=0}^{T} \gamma^t \, r(s_t, a_t)\right] \tag{15}$$

$\gamma \in (0,1)$ represents the discount factor. According to the policy gradient theorem [36], the gradient of the objective function can be expressed as:

$$\nabla_\theta J(\theta) = \mathbb{E}_{\tau \sim \pi_\theta}\left[\sum_{t=0}^{T} \nabla_\theta \log\pi_\theta(a_t \mid s_t) \cdot A^\pi(s_t, a_t)\right] \tag{16}$$

where $A^\pi(s,a) = Q^\pi(s,a) - V^\pi(s)$ denotes the advantage function.

To enhance sample efficiency, PPO implements importance sampling techniques, reformulating the objective function as:

$$J(\theta) = \mathbb{E}_{\tau \sim \pi_{\theta \text{ old}}}\left[\frac{\pi_\theta(a \mid s)}{\pi_{\theta \text{ old}}(a \mid s)} A^{\pi_{\theta \text{ old}}}(s,a)\right] \tag{17}$$

where $\pi_{\theta_{old}}$ represents the old policy, and $\pi_\theta$ represents the new policy.

To prevent excessively large policy updates that could destabilize training, PPO incorporates a clipping mechanism, formulating the following objective function:

$$L^{CLIP}(\theta) = \mathbb{E}_t[\min(r_t(\theta)A_t, \text{clip}(r_t(\theta), 1-\epsilon, 1+\epsilon)A_t)] \tag{18}$$

where $r_t(\theta) = \frac{\pi_\theta(a_t|s_t)}{\pi_{\theta_{old}}(a_t|s_t)}$ denotes the policy probability ratio, and $\text{clip}(r, 1-\epsilon, 1+\epsilon)$ constrains $r_t$ within the interval $[1-\epsilon, 1+\epsilon]$, with $\epsilon$ representing the clipping threshold (typically valued between 0.1 and 0.3).

To reduce the variance in advantage function estimation, PPO employs the Generalized Advantage Estimation (GAE) method [37]:

$$\hat{A}_t = \sum_{l=0}^{T-t} (\gamma\lambda)^l \, \delta_{t+l}$$
$$\delta_t = r_t + \gamma V_\phi(s_{t+1}) - V_\phi(s_t) \tag{19}$$

where the smoothing coefficient $\lambda \in [0,1]$ regulates the weights of multi-step returns through exponential decay, establishing a balance between bias and variance, while $V_\phi$ represents the state value function.

The final optimization objective of PPO is as follows:

$$L^{\text{Total}}(\theta)=L^{CLIP}(\theta)-c_1L^{VF}(\theta)+c_2L^{\text{Entropy}}(\theta) \tag{20}$$

where the value function loss is:

$$L^{VF}(\theta)=\frac{1}{2}\left(V_\phi(s_t)-V_{\text{target}}(s_t)\right)^2$$

and $V_{target}$ is calculated through Monte Carlo returns or temporal difference TD($\lambda$).

The entropy regularization term is:

$$L^{\text{Entropy}}(\theta)=-\sum_a \pi_\theta(a \mid s_t)\log\pi_\theta(a \mid s_t)$$

This term serves to encourage policy exploration and prevent premature convergence to suboptimal solutions.

The coefficients $c_1, c_2$ in equation 20 control the relative weights of each loss component (typical values: $c_1=0.5, c_2=0.01$).

In our experimental framework, the DRL environment (including state space and reward function design) and the PPO algorithm implementation remain mutually independent. The innovative short-selling mechanism is implemented exclusively within the DRL environment module, without any modifications to the underlying PPO algorithmic logic. This modular design approach not only enhances the system's maintainability and extensibility, but more significantly, the short-selling mechanism proposed in this paper can be generalized to any portfolio optimization model incorporating short-selling strategies, thereby offering broader theoretical and practical value.

## 4.2 Neural Network Design

Early artificial neural networks encountered significant vanishing and exploding gradient problems when increasing network depth, which constrained the development of data-driven models. Breakthroughs in deep neural network technology, particularly innovations in gradient propagation and network optimization techniques, provided critical support for reinforcement learning algorithm development, leading to the emergence of DRL. Within the DRL framework, the design of appropriate neural network architectures is essential, as architecture performance directly influences agent learning efficiency and decision-making capabilities. Empirical evidence indicates that well-designed network structures substantially improve DRL performance in practical applications.

In our experimental phase, we evaluated several deep neural network architectures, including ResNet [38], VGG [39], and Vision Transformer (ViT) [40]. ResNet demonstrated superior performance in both training and backtesting scenarios; however, it required greater computational resources and exhibited slower training convergence. After conducting a comprehensive analysis of performance metrics versus computational efficiency, we selected VGG as our foundational network architecture and modified it by removing the dropout layers present in the original implementation. Fig.3 illustrates our proposed deep neural network architecture, which is specifically designed to accommodate the Actor-Critic structure of the PPO algorithm:

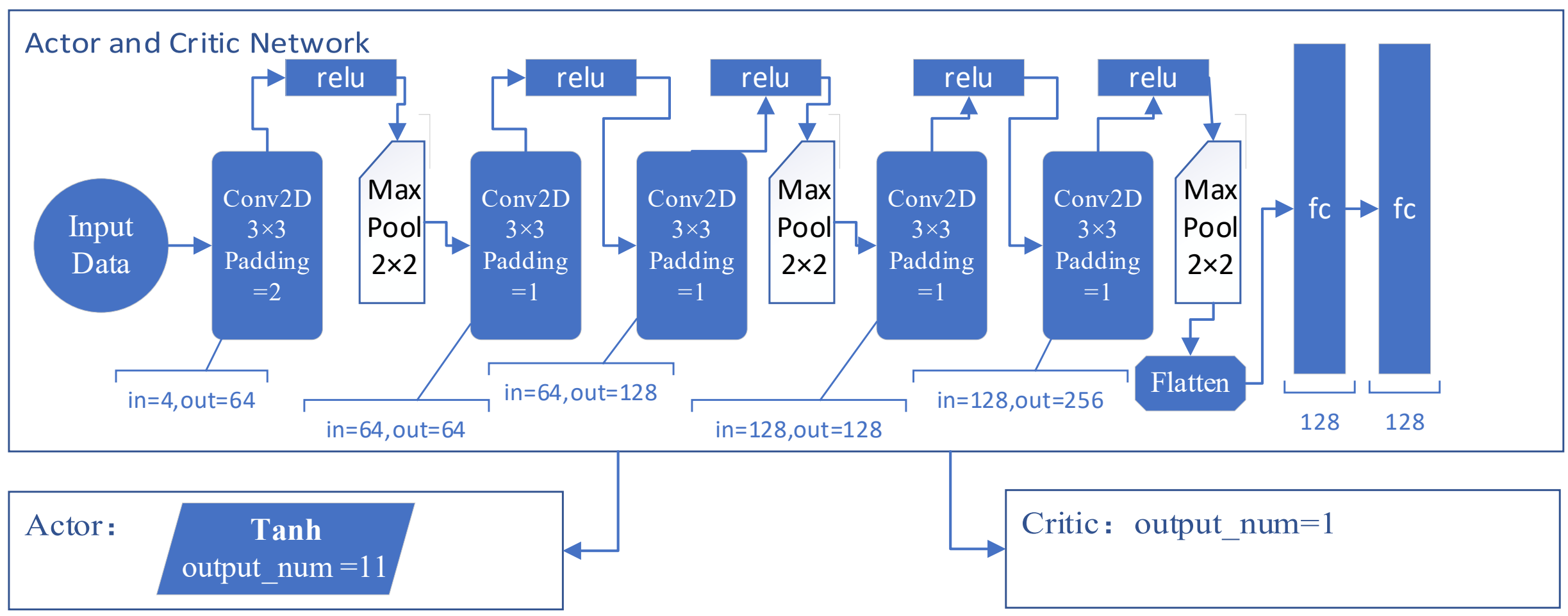

**Fig.3** Neural Network Structure

In Fig.3, "in" denotes the input channel dimensionality while "out" represents the output channel dimensionality. The architectural design comprises 5 sequential convolutional layers, each employing 3×3 convolutional kernels for feature extraction. The Max Pool layers implement maximum pooling operations to perform spatial downsampling of the extracted feature maps. Following feature processing through the final convolutional layer and subsequent maximum pooling, a Flatten operation transforms the multi-dimensional feature representations into a single-dimensional vector. This vector is then processed through two fully connected (fc) layers with 128 neurons each for linear transformation. At the network terminus, the Actor network incorporates a Tanh activation function to generate the action vector representing asset allocation weights, while the Critic network produces the value function estimation directly without activation transformation. The Tanh output dimension in the Actor network structure corresponds to a portfolio consisting of 11 assets (10 risky assets and 1 risk-free asset).

A critical implementation detail concerns the normalization procedure: since each asset weight is bounded within [-1, 1] due to the Tanh activation function, a normalization step is necessary to satisfy portfolio constraints. Specifically, we normalize the weights such that the sum of their absolute values equals 1, ensuring compliance with the portfolio constraint requirements specified in Formula 3.

## 5 EMPIRICAL TESTS

### 5.1 Data Selection, Preprocessing and Assumptions

The investment portfolio in this study was constructed through random sampling from the CSI 300 Index constituent stocks, with deliberate inclusion of securities exhibiting sideways volatility or declining price trajectories. This selection criterion was implemented to enhance the model's capability to identify and leverage downward trend characteristics when establishing short positions during the training process. Stock data is extracted from daily trading records in the Wind database, with all prices forward-adjusted for corporate actions. Transaction constraints limit each individual stock to a single trade per trading day.

This research deliberately implemented a random selection methodology for portfolio construction rather than employing optimization-based selection strategies derived from traditional investment theory. The rationale for this approach is predicated on the hypothesis that a truly effective DRL model should demonstrate adaptive decision-making capabilities across diverse portfolio compositions, not merely those optimized through conventional selection criteria. The subsequent robust performance observed during backtesting of these randomly constructed portfolios provides more rigorous validation of the model's decision-making efficacy and generalization capabilities. Furthermore, this methodology inherently evaluates the model's performance stability across heterogeneous market environments, highlighting a fundamental advantage of DRL as a data-driven approach: its capacity to autonomously adapt to varying market conditions without requiring manual security screening or selection heuristics.

Asset selection is constrained solely by the temporal criterion of listing prior to December 31, 2012. This criterion ensures each asset contains at least 10 years of historical data, encompassing multiple complete market cycles including bull and bear markets and diverse market conditions. This extensive historical dataset provides the DRL model with sufficient training samples to identify and learn various market characteristics. The research further assumes perfect liquidity for all risk assets, instantaneous trade execution, and zero market impact from trading activities.

### 5.2 Performance Metrics, Backtesting Period and Comparative Optimization Models

Performance evaluation metrics employed in this study include: annualized return E(R), annualized volatility Std(R), annualized Sharpe ratio (Sharpe), annualized Sortino ratio (Sortino), maximum drawdown (MDD), Calmar ratio (Calmar), win rate (%of+Ret), and average profit-to-loss ratio (Ave P/Ave L).

This research implements a backtesting horizon of approximately six months to evaluate the model's optimization capabilities. With an annual trading calendar of 252 days, six months equates to approximately 128 trading days. To mitigate overfitting risk while enhancing model generalization, we implemented the methodology proposed by Wassname [41] in their open-source GitHub repository. This approach involves random sampling of 128 consecutive trading days from the complete dataset to form individual training episodes. Consequently, the six-month backtesting horizon maintains methodological consistency with the model's episodic sampling protocol.

Fig.4 presents a visualization of both training and testing datasets. All price data displayed have undergone normalization processing, whereby each asset's price series is standardized by dividing all values by the opening price recorded on the final trading day. This normalization methodology facilitates clear visual comparison of price trends across portfolio components with inherently different price magnitudes.

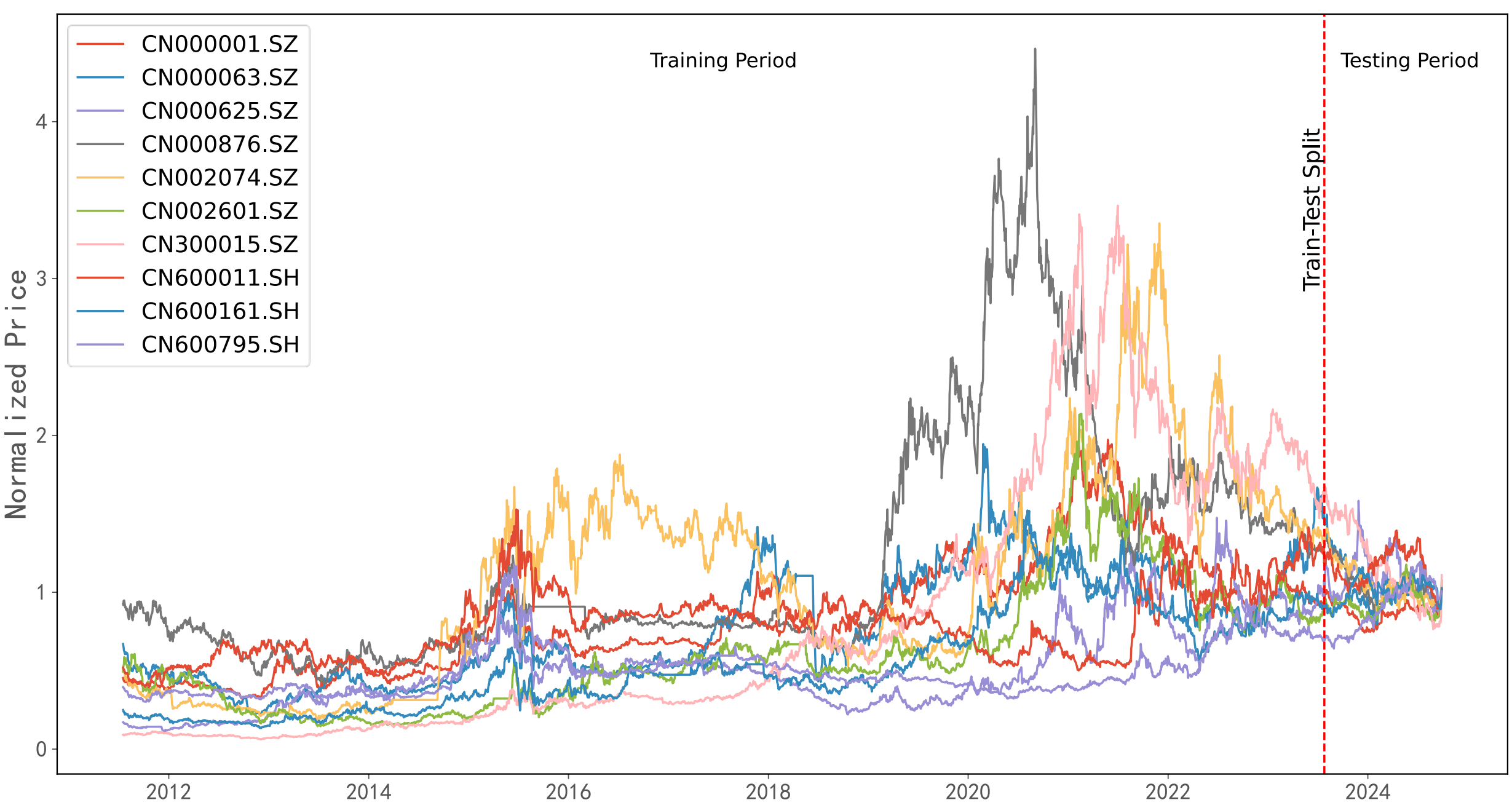

**Fig.4 Normalized Price Trends with Train&Test Split**

The backtesting interval comprises exclusively out-of-sample data that was rigorously segregated from the training dataset. The agent (functioning as the investment decision-maker) encounters these data sequences only during the evaluation phase, with no prior exposure to or information about future price trajectories. This strict separation ensures unbiased assessment of the model's predictive capabilities under realistic conditions. Table 1 provides comprehensive temporal specifications for both training and backtesting data periods:

**Table 1 Data Ranges for Training and Backtesting**

| Asset | Training Period | Testing Period |
|---|---|---|
| Stock Portfolio | 6/1/2011 - 7/20/2023 | 7/28/2023 - 2/2/2024<br>11/8/2023 - 5/22/2024 |

As evidenced in Table 1, our DRL model underwent comprehensive performance evaluation across two distinct backtesting periods to rigorously assess its optimization efficacy and robustness under diverse market regimes.

In the model comparison phase, this research benchmarks the DRL model against multiple established portfolio optimization frameworks. All comparative models utilize the Riskfolio-lib asset optimization package with default parameter configurations, and asset returns are uniformly calculated from closing prices. The benchmark optimization methodologies include: Classic Mean Variance (MV), Conditional Value at Risk (CVaR), Entropic Value at Risk (EVaR), Risk Parity (RP), Hierarchical Risk Parity (HRP), Hierarchical Equal Risk Contribution (HERC), and Nested Clustered Optimization (NCO). Several of these models incorporate various optimization objectives including risk minimization (MinRisk), Sharpe ratio maximization (Sharpe), utility function maximization (Utility), and return maximization (MaxRet). Due to the inherent subjectivity in utility function specification and the suboptimal backtesting performance exhibited by both utility function maximization and return maximization in preliminary experiments, the comparative analysis restricts optimization objectives exclusively to risk minimization and Sharpe ratio maximization.

For the comparative analysis of these traditional optimization models, this paper employs a 1-year historical data window (252 trading days) and implements a rolling window methodology for asset weight forecasting, thereby addressing the inherent limitation of traditional econometric optimization models that fail to capture dynamic weight evolution. Specifically, for the weight prediction on September 1, 2021, these models analyze the preceding 1-year historical data through August 31, 2021, with this process advancing sequentially until the complete backtesting interval is covered. Transaction costs incurred from portfolio rebalancing are calculated using formula 11, with transaction cost rates ($\mu_t = 0.0015$) for all assets maintained consistently with those specified in the DRL model.

5.3 Training Results and Robustness Testing

Deep Reinforcement Learning (DRL) represents a novel sequential statistical decision-making methodology that utilizes neural networks to model and estimate conditional probability distributions and expected returns across state-action spaces. At each time step, the agent conducts online statistical inference based on current observations while concurrently optimizing its decision policy through systematic exploration and experience accumulation. This process implements iterative statistical learning designed to maximize expected cumulative rewards. DRL integrates the powerful function approximation capabilities of deep learning with the sequential decision framework of reinforcement learning, establishing an end-to-end approach for statistical modeling and optimization.

Traditional econometric testing methods founded on linear assumptions prove inadequate for effectively evaluating the statistical significance of DRL models, whereas the convergence of training rewards offers a more appropriate evaluation criterion. Convergent behavior demonstrates the agent's capacity to generate consistent profits in historical market environments, constituting a necessary condition for model stability and robustness. This necessity derives from both the non-linear characteristics of DRL and its adaptive learning mechanisms; a converged reward function signifies that the model has captured stable patterns rather than overfitting to market noise. Moreover, reward convergence indicates the agent has developed a generalizable strategy capable of maintaining consistent performance across varied market conditions within the training distribution.

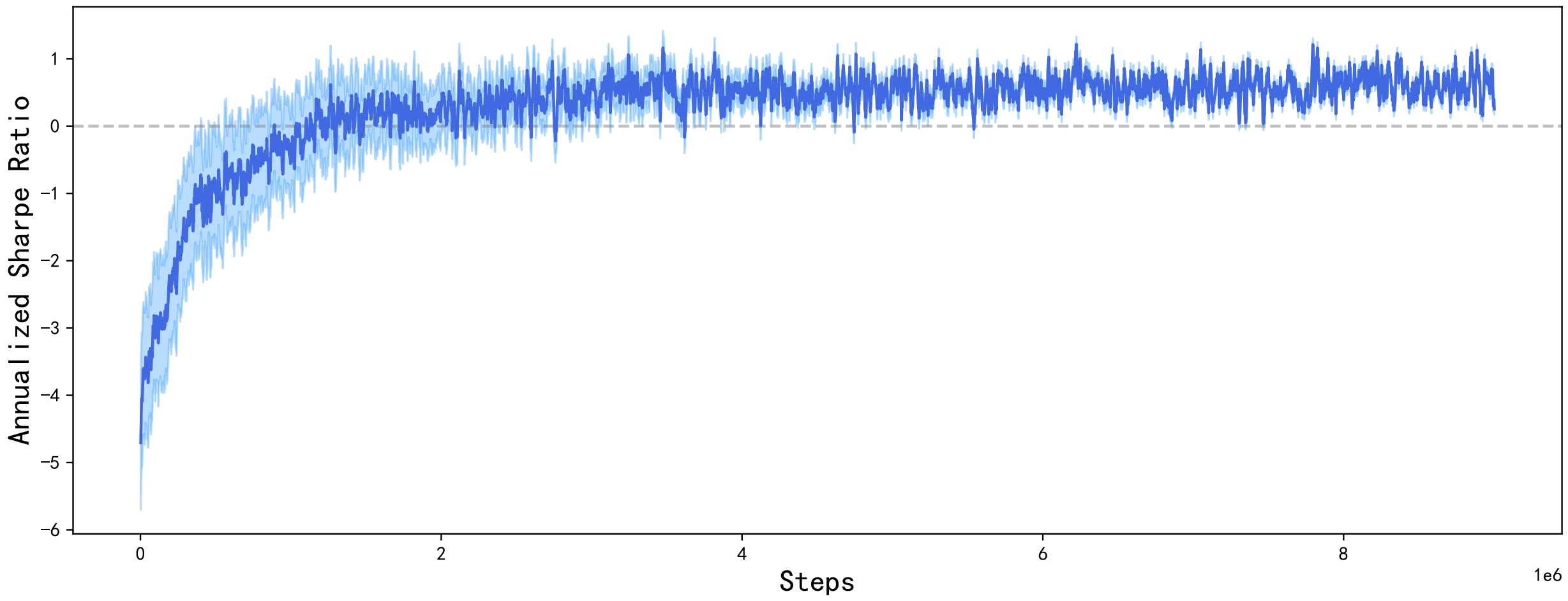

**Fig.5** Training rewards

As illustrated in Fig.5, the horizontal axis represents the number of training steps, while the vertical axis denotes the average Sharpe ratio. Throughout the 9 million training steps, the reward value demonstrates a distinct positive correlation with the training progression, indicating continuous optimization of agent performance. It should be noted that the training results exhibit certain volatility due to the implementation of a random seed mechanism without fixed values. Achieving a model with stable convergence and superior backtesting performance typically necessitates multiple training iterations to identify optimal configurations. This multi-round validation approach enhances model reliability and robustness. Throughout the training duration, reward values exhibit significant convergence characteristics, with annualized Sharpe ratios stabilizing within the 0.1 to 0.9 range and training rewards consistently converging above zero. These findings confirm the agent's sustained capacity to generate returns within known environments, substantiating the robustness of the trained model.

## 5.4 Backtesting results

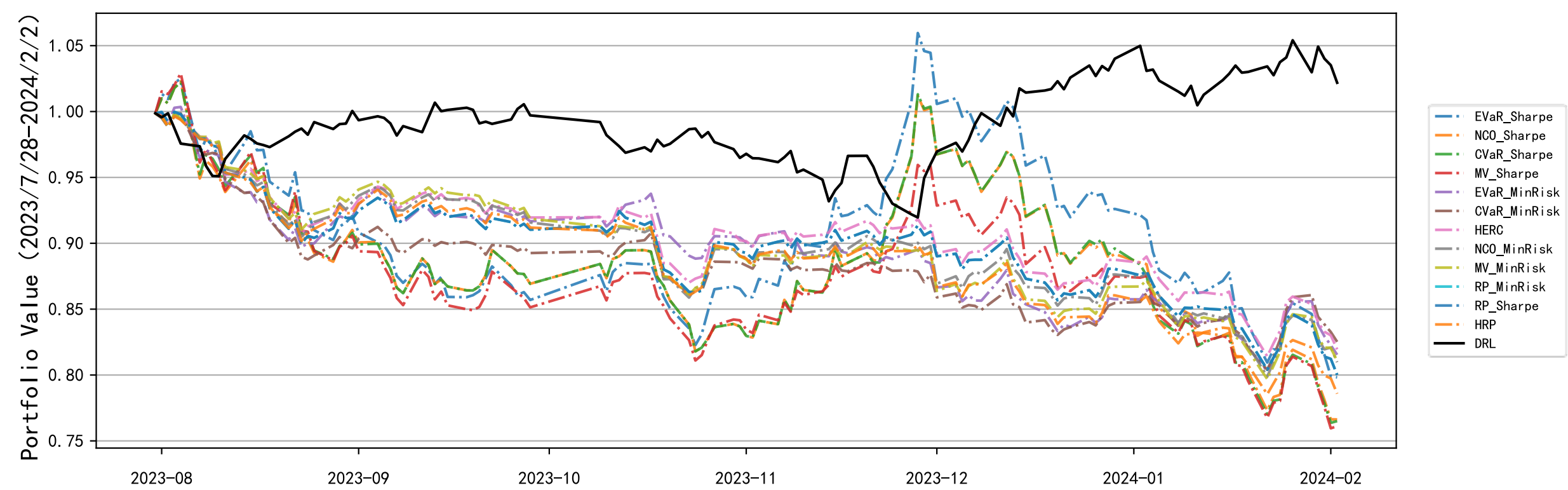

**Fig.6 Comparison of Optimized Portfolio Performance（2023/7/28-2024/2/2）**

Fig.6 illustrates that the DRL portfolio optimization model incorporating a short-selling mechanism exhibits robust growth throughout the backtesting period spanning July 28, 2023, to February 2, 2024. With the exception of a temporary correction between early October and late November 2023, the model maintains a consistent upward trajectory. Conversely, the traditional optimization model demonstrates weak performance across the entire backtesting period; despite experiencing a brief recovery from mid-November to December 2023, it subsequently resumes its downward trend. Additionally, while traditional optimization models display similar trend patterns, the DRL optimization model charts a distinctly different performance trajectory.

**Table 2** Results of optimization models for stocks (2023/7/28-2024/2/2)

| | E(R) | Std(R) | Sharpe | Sortino | MDD | Calmar | %of+Ret | Ave.P / Ave.L |
|---|---|---|---|---|---|---|---|---|
| DRL | 0.0422 | 0.1407 | 0.3000 | 0.5711 | 8.65% | 0.4881 | 0.4883 | 1.0816 |
| MV-MinRisk | -0.4134 | 0.1448 | -2.8542 | 3.7238 | 20.01% | -2.0475 | 0.4531 | 0.7395 |
| MV-Sharpe | -0.5372 | 0.2428 | -2.2124 | -3.0618 | 26.18% | -2.0515 | 0.4453 | 0.8605 |
| CVaR-MinRisk | -0.3782 | 0.1403 | 2.6952 | -3.5019 | 19.49% | -1.9405 | 0.4375 | 0.8072 |
| CVaR-Sharpe | -0.5272 | 0.2593 | -2.0331 | -3.0180 | 25.28% | -2.0849 | 0.4531 | 0.8555 |
| EVaR-MinRisk | -0.4018 | 0.1535 | -2.6160 | -3.5374 | 19.97% | -2.0114 | 0.4140 | 0.9059 |
| EVaR-Sharpe | -0.4437 | 0.2563 | -1.7309 | -2.7805 | 24.66% | -1.7986 | 0.4296 | 0.9954 |
| RP-MinRisk | -0.4391 | 0.1508 | -2.9110 | -4.1504 | 19.97% | -2.1986 | 0.4296 | 0.8214 |
| RP-Sharpe | -0.4391 | 0.1508 | -2.9110 | -4.1504 | 19.97% | -2.1986 | 0.4296 | 0.8214 |
| HRP | -0.4744 | 0.1472 | -3.2216 | -4.4430 | 21.33% | -2.2237 | 0.4375 | 0.7510 |
| HERC | -0.3920 | 0.1454 | -2.6953 | -3.7193 | 18.49% | 2.1196 | 0.4687 | 0.7216 |
| NCO-MinRisk | -0.4155 | 0.1466 | -2.8339 | -3.6801 | 20.00% | -2.0775 | 0.4296 | 0.8169 |
| NCO-Sharpe | -0.5240 | 0.2607 | -2.010 | -2.8908 | 25.08% | -2.0888 | 0.4296 | 0.9407 |

Table 2 reveals that during the backtesting period from July 28, 2023, to February 2, 2024, the DRL model incorporating a short-selling mechanism exhibits substantial advantages. The DRL model achieved a positive annualized return of 4.22%, whereas all traditional models incurred significant losses, with the MV-Sharpe model performing most poorly with losses reaching 53.72%. With respect to risk management, the DRL model demonstrated superior performance, maintaining a maximum drawdown of merely 8.65%, substantially below the drawdown ranges of 18.49% to 26.18% observed in traditional models. From a risk-adjusted return perspective, the DRL model generated a Sharpe ratio of 0.30 and a Sortino ratio of 0.57, distinguishing itself as the only model to produce positive values. By contrast, traditional models generally exhibited Sharpe ratios between -2 and -3, with the HRP model recording the lowest value at -3.22. The DRL model further excelled in trading efficiency metrics, achieving the highest win rate at 48.83% and the highest profit-to-loss ratio at 1.08 among all models evaluated. Notably, DRL stands as the only model with a profit-to-loss ratio exceeding 1, indicating that its average gains from profitable trades surpass its average losses from unprofitable positions. Comprehensively assessed across profitability, risk management, and trading efficiency dimensions, the DRL model substantially outperforms conventional portfolio optimization methodologies.

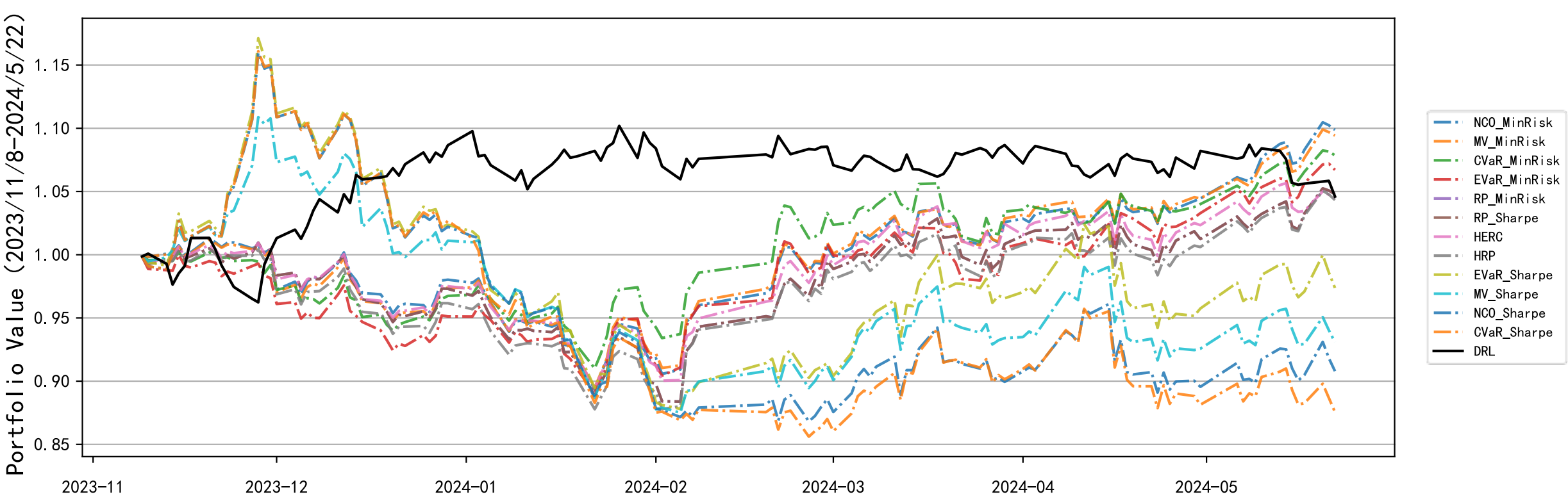

**Fig.7** Comparison of Optimized Portfolio Performance（2023/11/8-2024/5/22）

Fig.7 demonstrates that the DRL portfolio optimization model incorporating a short-selling mechanism exhibits robust performance during the backtesting period spanning November 8, 2023, to May 22, 2024. The portfolio appreciates consistently from early November 2023 through January 2024, subsequently maintaining this valuation level until the conclusion of the investment horizon. In contrast, traditional optimization models manifest pronounced volatility throughout the entire backtesting period, characterized by substantial drawdowns. While certain traditional models—notably MV-MinRisk, CVaR-MinRisk, and HERC—ultimately generated positive returns, all exhibited characteristic homogeneity in their optimization capabilities.

**Table 3** Results of optimization models for stocks（2023/11/8-2024/5/22）

| | **E(R)** | **Std(R)** | **Sharpe** | **Sortino** | **MDD** | **Calmar** | **%of+Ret** | **Ave. P / Ave. L** |
|---|---|---|---|---|---|---|---|---|
| DRL | 0.0879 | 0.1420 | 0.6186 | 1.1750 | 5.07% | 1.7332 | 0.5038 | 1.0676 |
| MV-MinRisk | 0.1773 | 0.1657 | 1.0695 | 1.5839 | 11.53% | 1.5372 | 0.5390 | 1.0269 |
| MV-Sharpe | -0.1410 | 0.2420 | -0.5827 | -0.8297 | 21.08% | -0.6692 | 0.5390 | 0.7757 |
| CVaR-MinRisk | 0.1489 | 0.1628 | 0.9144 | 1.5077 | 9.18% | 1.6223 | 0.5156 | 1.0971 |
| CVaR-Sharpe | -0.2617 | 0.2570 | -1.0183 | -1.6519 | 26.26% | -0.9965 | 0.4921 | 0.8726 |
| EVaR-MinRisk | 0.1278 | 0.1719 | 0.7435 | 1.2961 | 10.92% | 1.1705 | 0.4765 | 1.2444 |
| EVaR-Sharpe | -0.0524 | 0.2668 | -0.1964 | -0.3160 | 25.02% | -0.2094 | 0.5078 | 0.9380 |
| RP-MinRisk | 0.0920 | 0.1729 | 0.5323 | 0.8894 | 12.44% | 0.7397 | 0.5234 | 0.9934 |
| RP-Sharpe | 0.0920 | 0.1729 | 0.5323 | 0.8894 | 12.44% | 0.7397 | 0.5234 | 0.9934 |
| HRP | 0.0833 | 0.1733 | 0.4808 | 0.7692 | 12.93% | 0.6619 | 0.5390 | 0.9278 |
| HERC | 0.0864 | 0.1647 | 0.5245 | 0.8475 | 11.35% | 0.7610 | 0.5468 | 0.9028 |
| NCO-MinRisk | 0.1854 | 0.1668 | 1.1112 | 1.6441 | 11.78% | 1.5738 | 0.5390 | 1.0333 |

| NCO-Sharpe | -0.1901 | 0.2573 | -0.7387 | -1.1444 | 25.14% | -0.7561 | 0.4843 | 0.9409 |
|---|---|---|---|---|---|---|---|---|

Table 3 demonstrates that during the backtesting period from November 8, 2023 to May 22, 2024, the DRL model with integrated short-selling mechanism achieved a Calmar ratio of 1.73, the highest among all tested models. This indicates the DRL model's capacity to generate optimal returns per unit of maximum drawdown risk. The DRL model recorded a maximum drawdown of only 5.07%, substantially lower than traditional models which exhibited drawdowns ranging from 9.18% to 26.26%. While the DRL model produced an annualized return of 8.79%, compared to 17.73% and 18.54% for MV-MinRisk and NCO-MinRisk respectively, it maintained favorable risk-adjusted metrics with a Sharpe ratio of 0.62 and a Sortino ratio of 1.18. Significantly, all models optimized for maximum Sharpe ratio—MV-Sharpe, CVaR-Sharpe, EVaR-Sharpe, and NCO-Sharpe—resulted in negative returns, with CVaR-Sharpe showing the largest annualized loss at 26.17%. From an operational perspective, the DRL model attained a 50.38% win rate and a profit-loss ratio of 1.07, reflecting its capability to balance return generation with risk management. Although certain traditional approaches delivered higher absolute returns during the test period, the DRL model's combination of optimal Calmar ratio, minimal drawdown, and consistent performance illustrates its comprehensive effectiveness in portfolio management.

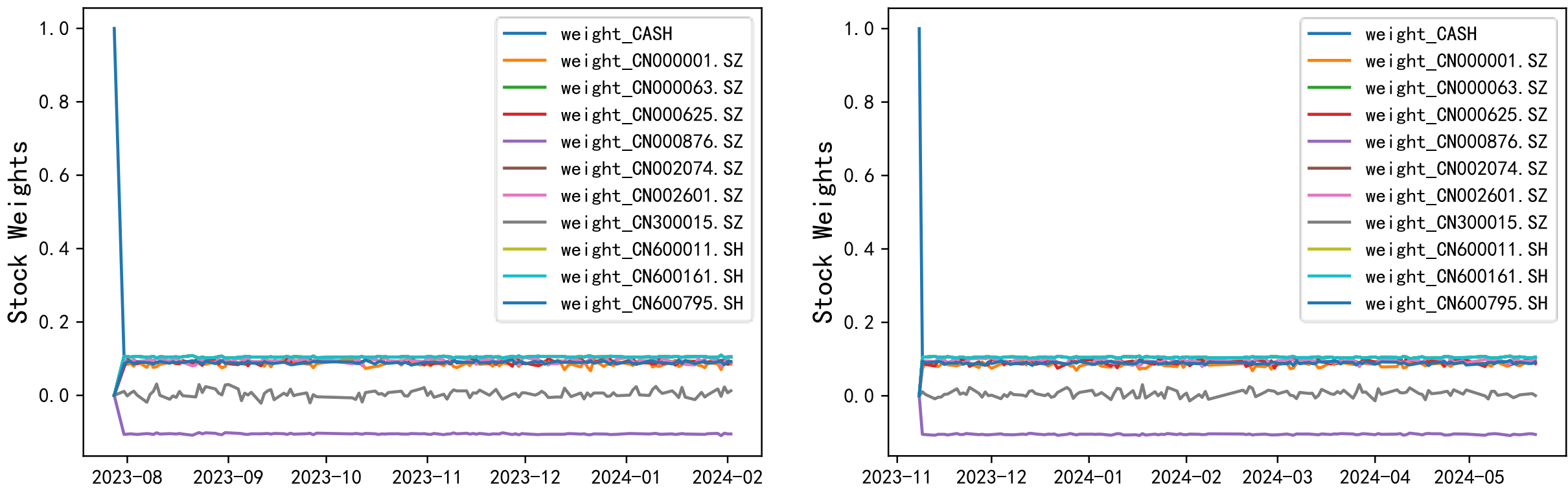

**Fig.8** Portfolio Weight Allocation of DRL Model with Short-Selling

The asset weight allocation analysis presented in Fig.8 indicates that the DRL optimization model demonstrates a systematic risk management approach. For the security 000001.SZ, the model allocates near-zero weights across both backtesting periods, while implementing a relatively uniform allocation strategy for the remaining assets. The model assigns approximately 0.1 weight to cash holdings and establishes a short position of approximately -0.1 for 000876.SZ. This allocation framework suggests the model has developed a disciplined investment methodology through its training process: systematically underweighting historically underperforming assets while distributing capital (including cash reserves) with relative uniformity across the remaining investment universe. This allocation strategy contributes to enhanced portfolio risk mitigation characteristics.

## 6 CONCLUSIONS

This study develops a long-short portfolio optimization framework utilizing deep reinforcement learning, exploring effective methodologies for generating excess returns in the A-share market through a short-selling mechanism aligned with practical trading constraints. The research yields three principal findings:

First, through systematic analysis of short-selling mechanics, we developed an innovative framework that constrains the sum of absolute asset weight values to 1. This approach overcomes limitations in traditional short-selling weight calculation methodologies during continuous trading scenarios and more accurately captures return dynamics in practical trading environments, establishing a robust theoretical foundation for long-short portfolio optimization.

Second, the DRL model with integrated short-selling functionality demonstrates significant performance enhancement. Across two distinct backtesting periods, the model generated consistent positive returns while maintaining substantially lower maximum drawdowns compared to traditional optimization approaches. During the July 2023 to February 2024 period, the DRL model achieved an annualized return of 4.22% when traditional models predominantly incurred substantial losses. In the November 2023 to May 2024 interval, the model generated an 8.79% annualized return while maintaining a maximum drawdown of only 5.07%.

Third, the asset allocation profile indicates the DRL model developed a systematic investment methodology. The model consistently underweights or avoids assets with suboptimal historical performance while maintaining appropriate cash reserves and implementing measured short positions, reflecting disciplined risk management principles. This balanced allocation approach effectively enhances the portfolio's risk-adjusted performance metrics.

This study presents several limitations. While we randomly selected stocks exhibiting sideways or downward movements from the CSI 300 index for our experiments, these securities nevertheless demonstrate overall upward trajectories when examined across extended historical periods. This characteristic constrains the DRL model's capacity to fully capture downward trends and associated trading signals during the training process, despite the incorporation of short-selling mechanisms. Relative to long-only frameworks, optimization capability during bullish market conditions remains an area for potential enhancement. Consequently, when utilizing daily price data, the proposed long-short portfolio optimization model demonstrates greater efficacy for securities exhibiting historical price consolidation or declining trends rather than assets characterized by substantial upward momentum.

Future research directions include further refinement of the DRL environment through integration of intraday data (1-minute, 5-minute intervals) in conjunction with the T+1 trading constraints specific to China's A-share market. Higher frequency data would facilitate more precise modeling of asset trading characteristics. As artificial intelligence technologies continue to integrate with financial applications, the implementation potential of DRL in portfolio optimization will expand. The convergence of DRL with other emerging technologies promises significant advancements in quantitative investment methodologies and accelerated development of intelligent investment decision systems.